\documentclass[prb,showpacs,twocolumn,aps,superscriptaddress,a4paper]{revtex4-1}
\usepackage{dcolumn,amssymb,amsmath,amsfonts,graphicx,latexsym,color}
\usepackage{epstopdf}
\usepackage{subfigure}

\begin{document}

\title{Topological phase transition in the quasiperiodic disordered Su-Schriffer-Heeger chain}

\author{Tong Liu}
\affiliation{Department of Physics, Southeast University, Nanjing 211189, China}
\author{Hao Guo}
\thanks{Corresponding author: guohao.ph@seu.edu.cn}
\affiliation{Department of Physics, Southeast University, Nanjing 211189, China}

\date{\today}

\begin{abstract}
We study the stability of the topological phase in one-dimensional Su-Schrieffer-Heeger chain subject to the quasiperiodic hopping disorder. We investigate two different hopping disorder configurations, one is the Aubry-Andr\'{e} quasiperiodic disorder without mobility edges and the other is the slowly varying quasiperiodic disorder with mobility edges. With the increment of the quasiperiodic disorder strength, the topological phase of the system transitions to a topologically trivial phase. Interestingly, we find the occurrence of the topological phase transition at the critical disorder strength which has an exact linear relation with the dimerization strength for both disorder configurations. We further investigate the localized property of the Su-Schrieffer-Heeger chain with the slowly varying quasiperiodic disorder, and identify that there exist mobility edges in the spectrum when the dimerization strength is unequal to 1. These interesting features of models will shed light on the study of interplay between topological and disordered systems.
\end{abstract}

\maketitle

\section{Introduction}
\label{n1}
Topological insulators~\cite{hasan2010,moore2010,fu2007,kitaev2009} (TIs), a unique class of electronic materials, have not only a bulk gap but also symmetry-protected gapless states localized at the sample boundaries. Topological features of TIs are expected to be immune to perturbations of the fluctuation and the environmental noise, and thereby have great potential applications in quantum information processing. Since the effect of disorder is inevitable in real materials, physical properties of topological systems in the presence of disorder have drawn considerable attention in the past decades~\cite{wray2011,ran2009,beidenkopf2011,nomura2011}. The combination of topology and disorder can induce rich novel quantum phenomenons. For weak disorder, the robustness of
topological states has been demonstrated for various topological systems, especially the quantum spin Hall states~\cite{yu2010}. With the increase of the disorder strength, the transition from topological phase to topologically trivial phase can occur. Besides destroying topological states, the medium strength disorder can induce the so called topological Anderson insulator~\cite{li2009,groth2009,jiang2009,song2012,guo2010} (TAI), i.e., a topologically trivial state is driven into a topological state by disorder.

However, the direct observation of the influence of disorder on TIs is difficult due to the difficulty to precisely control the disorder in solid-state experiments. Benefited from the development in the manipulation of ultracold atoms, exploring both disorder and topology via the quantum simulation in artificial systems has become exercisable. In one-dimensional (1D) system the topological/trivial
feature of an insulator is completely determined by the presence/absence of the chiral symmetry. The Su-Schrieffer-Heeger (SSH) model~\cite{su1979,mondragon2014,li2014,zhu2014} is one of the most simple and widely studied models belonging to the BDI symmetry class, which hosts two topologically distinguishable phases characterized by the winding number. Using a combination of Bloch oscillations and Ramsey interferometry, Ref.~\cite{atala2013} measures the Zak phase to study the topological feature of Bloch bands in a dimerized optical lattice described by 1D SSH model. A very recent work~\cite{meier2018}, in which a 1D chiral symmetric SSH chain with controllable off-diagonal (hopping) disorder is synthesized based on the laser-driven ultracold atoms~\cite{an2017}, systematically explores the disorder effect on topology. They observe the robustness of topological state immune to weak random hopping disorder, and also the topologically non-trivial to trivial transition
at very strong hopping disorder. More interestingly, they find the TAI phase in which the band structure of a topologically trivial chain is driven to a non-trivial one by adding the random hopping disorder.

On the other hand, the bichromatic optical lattices~\cite{modugno2009,schreiber,d2014} with incommensurate wavelengths have also attracted enormous attention in the study of Anderson localization. A notable work experimentally investigates the localization property
of 1D Aubry-Andr\'{e} (AA) model~\cite{aubry} by use of ultracold atoms in the incommensurate/quasiperiodic optical lattice~\cite{roati2008}. The AA model has a self-dual symmetry and can undergo a sharp localization transition at the self-dual point without mobility edges. There also exists a class of systems with slowly varying quasiperiodic disorder which can host mobility edges~\cite{sarma1988,sarma1990}. Motivated by the highly precise control in these ultracold atomic experiments, an interesting
question can be raised: what is the fate of the topological non-trivial state when
the quasiperiodic hopping disorder is introduced in 1D SSH chain, besides the random hopping disorder? To examine this question, we propose two quasiperiodic disordered SSH models with two different disorder configurations and investigate how the topological phase transition occurs.

The rest of the paper is organized as follows. In Sec.~\ref{n2}, we investigate the topological phase transition in 1D SSH chain by adding the AA quasiperiodic disorder. In Sec.~\ref{n3}, we investigate the topological phase transition in 1D SSH chain by adding the slowly varying quasiperiodic disorder. We also investigate the localized property (mobility edges) of this system besides the topological phase transition. The conclusion is summarized in Sec.~\ref{n4}.

\section{the AA quasiperiodic disorder}
\label{n2}

\begin{figure}
	\centering
	\includegraphics[width=0.5
	\textwidth]{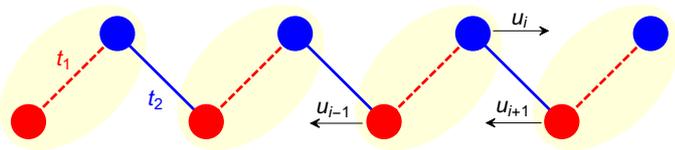}\\
	\caption{(Color online) Schematics of 1D SSH chain with the displacement $u_i$. The system consists
   of two sublattices indicated by red and blue filled circles respectively. Hopping amplitudes are staggered by $t_1$ (red dashed line) and $t_2$ (blue solid line).}
	\label{001}
\end{figure}
The SSH model describes a 1D chain of spinless fermions with alternating hopping strengths between the neighboring tight-binding lattices. For a chain with $L$ lattices and open boundary conditions, the Hamiltonian of the model is expressed as
\begin{equation}\label{ham1}
    \begin{split}
    \hat H=-\sum_{i}^{L}( t_{i,i+1} \hat{c}^\dag_{i+1}\hat{c}_{i} + H.c.),
    \end{split}
\end{equation}
where $\hat{c}^\dagger_i$ ($\hat{c}_i$) is the fermion creation
(annihilation) operator of the $i$-th lattice, $t_{i,i+1}=t+a(u_{i+1}-u_i)$ is the nearest-neighbor
hopping amplitude, $a$ is the displacement coupling constant and $u_i$ is the configuration coordinate for the displacement of the $i$-th lattice. In the ideal limit, the displacement is of the perfectly periodic form, $u_{i+1}-u_i=(-1)^i \frac{\lambda}{a}$. Thus, the nearest-neighbor
hopping amplitude becomes $t_{i,i+1}=t+(-1)^i\lambda$, where the hopping unit $t$ and the dimerization strength $\lambda$ are both constants. This is the ordinary SSH model. When $\lambda>0$ the system is in the topological phase with the presence of twofold-degenerate zero-energy edge states at two ends, whereas when $\lambda<0$ the system is in the topologically trivial phase without the presence of zero-energy edge states.

When the displacement becomes disordered, i.e., $u_{i+1}-u_i=(-1)^i \frac{\lambda+\delta_{i}}{a}$, Eq.~(\ref{ham1}) can be written as
\begin{equation}\label{ham2}
    \begin{split}
    \hat H=-\sum_{i}^{\frac{L}{2}}( t_1 \hat{c}^\dag_{2i-1}\hat{c}_{2i} + H.c.)
    -\sum_{i}^{\frac{L}{2}-1}( t_2 \hat{c}^\dag_{2i}\hat{c}_{2i+1}+H.c.),
    \end{split}
\end{equation}
where the intracell hopping $t_1= t- \lambda-\delta_{i} $ and the intercell hopping $t_2= t + \lambda+\delta_{i}$ with $\delta_{i}=\delta\cos(2\pi\beta{i}+\phi)$ represent the AA quasiperiodic disorder, see Fig.~\ref{001} for illustration. The total Hamiltonian still maintains the chiral symmetry. In this paper we concentrate on the stability of the topological phase, so we only focus on the situation with $\lambda>0$ hereafter. A typical choice of the parameters is $\delta>0$, $\beta=(\sqrt{5}-1)/2$ and $\phi = 0$. For convenience, $t = 1$ is set as the energy unit.

\begin{figure}
	\centering
	\includegraphics[width=0.5
	\textwidth]{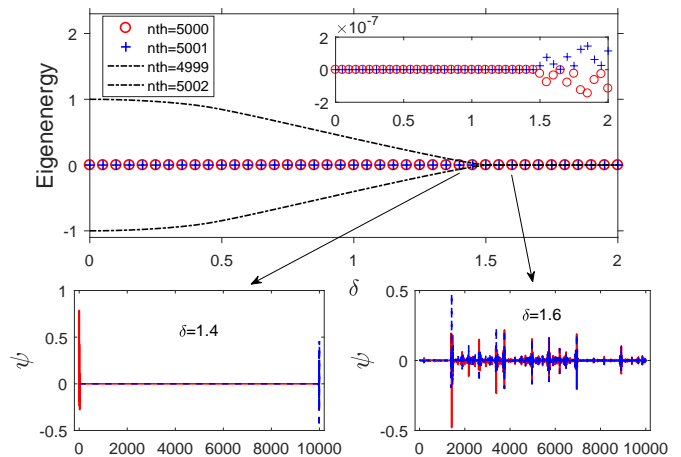}\\
	\caption{(Color online) The spectrum (only the middle four eigenvalues are shown) of the Hamiltonian (\ref{ham2}) with $\lambda=0.5$ as a function of $\delta$ under the open boundary condition. The total number of sites is set as $L=10000$ hereafter. The $5000$th and $5001$th eigenvalues stay at zero until $\delta=1.5$, see the blow up of the inset. The spatial distributions of $\psi$ for the $5000$th and $5001$th eigenvalues with various $\delta$ are shown in the lower figures. The lower left picture corresponds to $\psi$ of zero-energy modes in the topological region, and the lower right picture corresponds to $\psi$ of nonzero-energy modes in the topologically trivial region.}
	\label{002}
\end{figure}

By numerically diagonalizing the Hamiltonian~(\ref{ham2}), we can get the eigenvalues and the wave-functions (denoted by $\psi$) of the system.
We show the spectrum when $\lambda=0.5$ under the open boundary condition in Fig.~\ref{001}, where there is a regime
with nonzero energy gaps in the range $\delta <1.5$ and there are zero-energy modes in the midgap.
As long as the chiral symmetry is preserved, zero-energy modes cannot be removed by the weak disorder. To show the difference between wave-functions with the zero-energy and nonzero-energy modes clearly, we plot the spatial distributions of wave-functions for the midmost excitations (the $5000$th and $5001$th eigenvalues) with different $\delta$'s. When $\delta=1.4$, the energy gap is still finite, and the wave-functions with zero-energy modes are located at the left (right) end of the chain and decay very quickly away from the left (right) edge with no overlap, as shown in Fig.~\ref{001}. However, when $\delta=1.6$, the energy gap is closed, the zero-energy modes disappear and the amplitudes of the wave-functions of the midmost excitations overlap together and are located within a finite range of the whole chain. Therefore, these results demonstrate that the system can undergo a transition from a topological phase to a topologically trivial phase when the strength of the quasiperiodic disorder $\delta$ exceeds a certain level.

\begin{figure}
  \centering
  \includegraphics[width=0.5
  \textwidth]{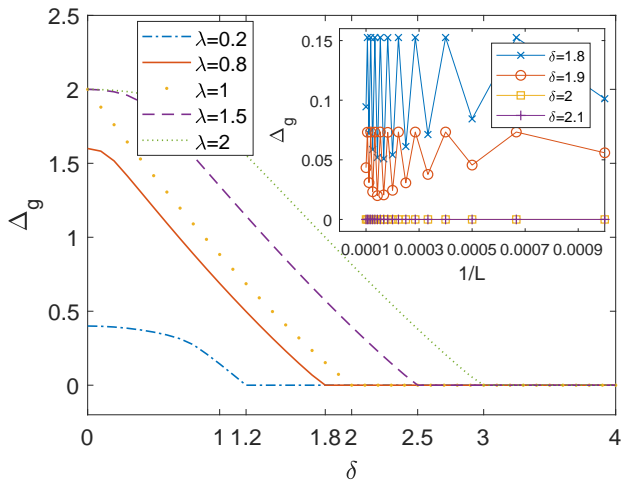}\\
  \caption{(Color online) $\Delta_g$ as a function of $\delta$ with various $\lambda$'s. The inset shows the finite size analysis of $\Delta_g$ near the critical point $\delta=1+\lambda$ ($\lambda=1$ as an example).}
  \label{003}
\end{figure}
We now wonder if there exists a fixed value of $\delta$ which denotes the gap-closing point~\cite{cai2013}. Due to the chiral symmetry, the eigenvalues appear in pairs. We arrange the eigenvalues in the ascending order and use $E'$ to denote the smallest eigenvalue which is larger than the zero energy. Thus, $\Delta_g=2E'$ can explicitly determine the gap-closing point and denote the topological phase boundary between different topological phases due to the bulk-boundary correspondence. In Fig.~\ref{003}, we plot the variation of energy gap $\Delta_g$ versus $\delta$ for different $\lambda$'s. It is clearly shown that there are finite gaps as $\delta$ is smaller than a critical value, i.e. $\delta<1+\lambda$, whereas the energy gaps vanish when $\delta>1+\lambda$. To visualize the result better we make a finite size analysis of the scaling behavior of $\Delta_g$ for $\lambda=1$ as a function of the inverse of system size in the inset of Fig.~\ref{003}. It exhibits an oscillating behavior at $\delta=1.8$, $1.9$,
which indicates that the energy gap is finite in
the regime with $\delta<1+\lambda$. Whereas the energy gap approaches zero at $\delta=2.0$, $2.1$, which indicates that $\delta=1+\lambda$ indeed denotes the gap-closing point. To strength the validity of our conclusion, we also systematically calculate the energy gap with other sets of $\delta$ and $\lambda$, and find the
similar behaviors.

\begin{figure}
	\centering
	\includegraphics[width=0.5
	\textwidth]{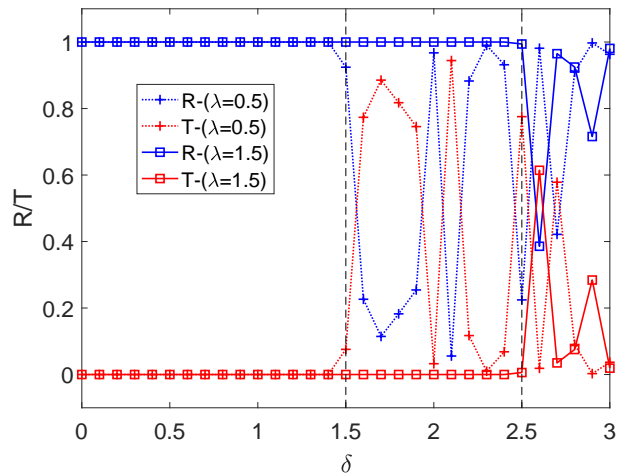}\\
	\caption{(Color online) The reflection probability $R$ and the transmission probability $T$ as a function of $\delta$ for systems with various $\lambda$'s.}
	\label{004}
\end{figure}

In general, the topological phase transition is characterized by the change of the topological invariant. Beyond the translational invariant system, it is more suitable to work with the transfer matrix~\cite{akhmerov2011,fulga2011,fulga2012,degottardi2013} to identify the topological phase of a finite disordered chain.
The transfer matrix formulation is designed to study the propagation of waves via a chain with arbitrary configurations. It thus offers a natural means of probing the localization nature of zero-energy modes and determining the topological characteristics of the SSH chain in the presence of the disorder.
From the tight-binding Hamiltonian~(\ref{ham1}) we can directly express the zero-energy transfer matrix $M_i$~\cite{fulga2011} in the site basis,
\begin{align}
&\left( \begin{array}{c}
         t_{i+1,i}\psi_i \\
	\psi_{i+1}
        \end{array}
 \right) = M_i \left( \begin{array}{c}
        t_{i,i-1}\psi_{i-1} \\
	\psi_{i}
        \end{array}
 \right),\label{M1}\\
&M_i = \left( \begin{array}{cc}
         0 & t_{i+1,i}  \\
	 - 1/t_{i+1,i} & 0
        \end{array}
 \right).
\label{M2}
\end{align}
Waves at the two ends of the chain ($L$ is even) are related by the total transfer matrix
\begin{equation}
 M =M_{L}M_{L-1}\cdots M_{2}M_{1}=\begin{pmatrix}
Z&0\\
0&1/Z
\end{pmatrix}.\label{M3}
\end{equation}
Then we transform to a new basis with right-moving and left-moving waves separated in the upper and lower two components by means of the unitary transformation $\Omega^{\rm T}\sigma_{y}\Omega^{\ast}=\sigma_{z}$ with $\sigma_{y}$ and $\sigma_{z}$ being the second and third Pauli matrices, and we get
\begin{equation}
{\cal M}=\Omega^{\rm T}M\Omega^{\ast}=\frac{1}{2Z}\begin{pmatrix}
Z^{2}+1&Z^{2}-1\\
Z^{2}-1&Z^{2}+1
\end{pmatrix},\label{M4}
\end{equation}
where
\begin{equation}
Z=  (-1)^{L/2}\, \prod_{i=1}^{L/2} \frac{t_{2i+1,2i}}{t_{2i,2i-1}} .\label{Z}
\end{equation}
We obtain the reflection amplitude $r$ and transmission amplitude $t$ from Eq~(\ref{M4})
\begin{equation}
r=\frac{1-Z^2}{1+Z^{2}},~~t=\frac{2Z}{1+Z^{2}}.\label{rt}
\end{equation}

Recalling Fig.~\ref{001}, the spatial distributions of the wave-functions of the midmost excitations for topologically distinct phases are completely different. The amplitudes associated with zero-energy modes in topological phase are exponentially localized at two ends of the chain with no overlap, whereas the amplitudes associated with topologically trivial phase overlap together and disperse within the whole chain. At the Fermi level (the zero energy), the transmission probability $T=tt^{\ast}$ via the chain must be zero because there are no probability distributions of particles from one end to the other in the topological phase. Hence the reflection probability $R=rr^{\ast}$ must be 1 due to the conservation of the probability current, i.e., a complete reflection process occurs. Whereas in topologically trivial phase both $T$ and $R$ of the Fermi level are finite and less than 1 due to finite probability distributions of particles within the whole chain.

Therefore, we can employ $T$ and $R$ as probes of the topological phase transition. In Fig.~\ref{004} we plot $T$ and $R$ of the midmost excitations
for $\lambda=0.5$ and $1.5$. As illustrated, $T\Rightarrow0$ and $R\Rightarrow1$ when the quasiperiodic disorder strength $\delta<1+\lambda$, whereas $0<T<1$ and $0<R<1$ when $\delta\geq1+\lambda$. The transfer matrix method is proved to be powerful and all numerical results are in excellent agreement with those by the finite-size analysis of the gap-closing point. Hence we can safely believe that the topological phase transition point is indeed located at $\delta=1+\lambda$.

In fact, the topological phase transition can be determined by the following mathematical conjecture. The limit of $Z$ in Eq.(\ref{Z}) is expressed as,
\begin{equation}
\lim\limits_{L\rightarrow\infty}Z= \lim\limits_{L\rightarrow\infty} (-1)^{L/2}\, \prod_{i=1}^{L/2} \frac{1+\lambda+\delta\cos(2\pi\beta{i})}{1-\lambda-\delta\cos(2\pi\beta{i})} .\label{Z1}
\end{equation}
It can be conjectured that if $\delta<1+\lambda$ Eq.~(\ref{Z1}) diverges and accordingly $R\Rightarrow1$ and $T\Rightarrow0$ in the large $L$ limit, whereas if $\delta\geq1+\lambda$ Eq.~(\ref{Z1}) converges and accordingly $0<R,T<1$. Although we can not prove this mathematical conjecture, we believe that this problem may bring some interests to both the physical and mathematical communities.

\section{the slowly varying quasiperiodic disorder}
\label{n3}
\begin{figure}
	\centering
	\includegraphics[width=0.5
	\textwidth]{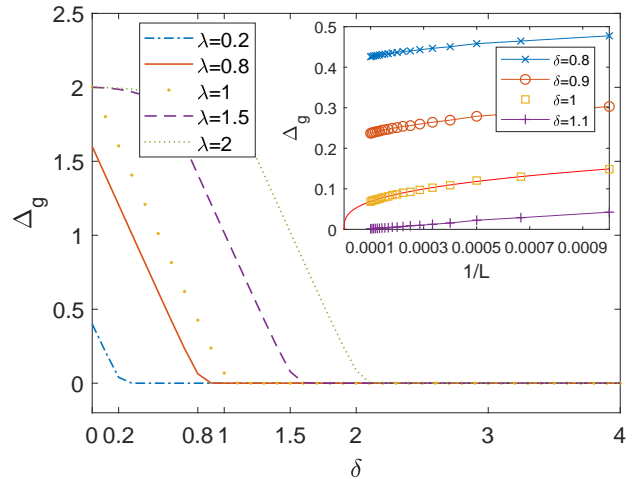}\\
	\caption{(Color online) $\Delta_g$ as a function of $\delta$ with various $\lambda$'s. The inset shows the finite size analysis of $\Delta_g$ near the critical point $\delta=\lambda$ ($\lambda=1$ as an example).}
	\label{005}
\end{figure}
\begin{figure}
	\centering
	\includegraphics[width=0.5
	\textwidth]{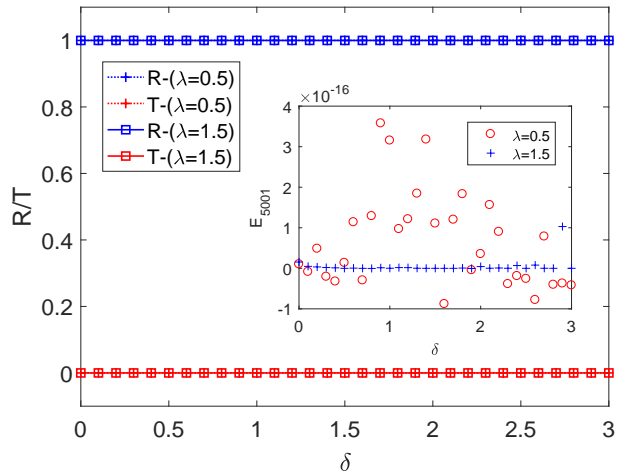}\\
	\caption{(Color online) The reflection probability $R$ and the transmission probability $T$ as a function of $\delta$ for systems with various $\lambda$'s. Surprisingly, the $5001$th eigenvalue with various $\lambda$'s stay at zero as $\delta$ increases, see the blow up in the inset.}
	\label{006}
\end{figure}

In this section we investigate the effect of the slowly varying quasiperiodic disorder on the topological phase of 1D SSH chain. The expression of Hamiltonian~(\ref{ham2}) remains unchange except the disorder term is replaced by $\delta_{i}=\delta\cos(2\pi\beta{i^{v}})$ with $0<v<1$. This type of disorder is revealed by Sarma et al.~\cite{sarma1988,sarma1990}, and can induce mobility edges located at the spectrum, unlike the AA disorder. Without loss of generality, we choose the parameter $v = 0.5$.

\begin{figure*}
    \begin{tabular}{cc}
    \subfigure[]{
    \begin{minipage}[t]{0.5\textwidth}
    \centering
    \includegraphics[width=1\textwidth]{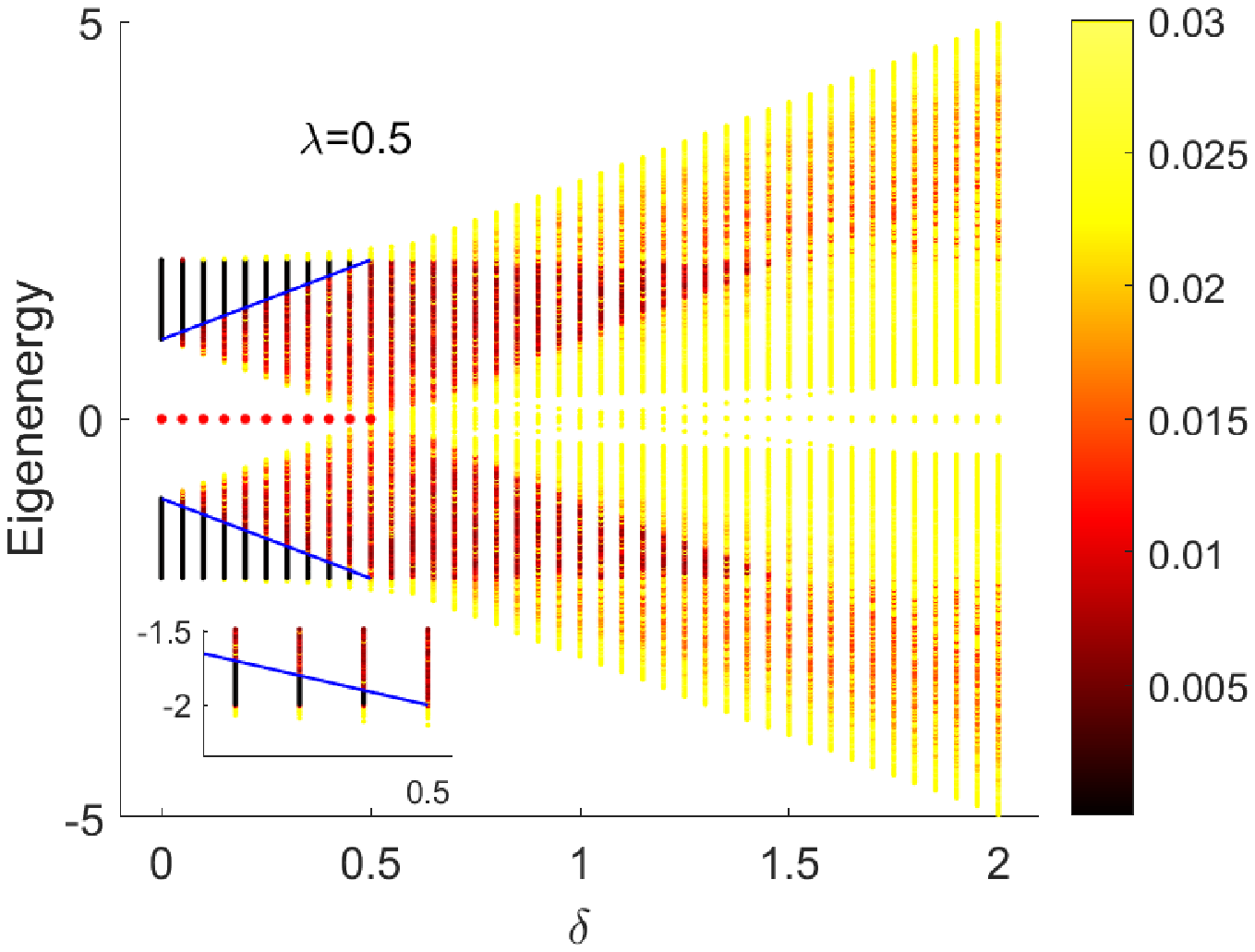}
    \end{minipage}}
    \subfigure[]{
    \begin{minipage}[t]{0.5\textwidth}
    \centering
    \includegraphics[width=1\textwidth]{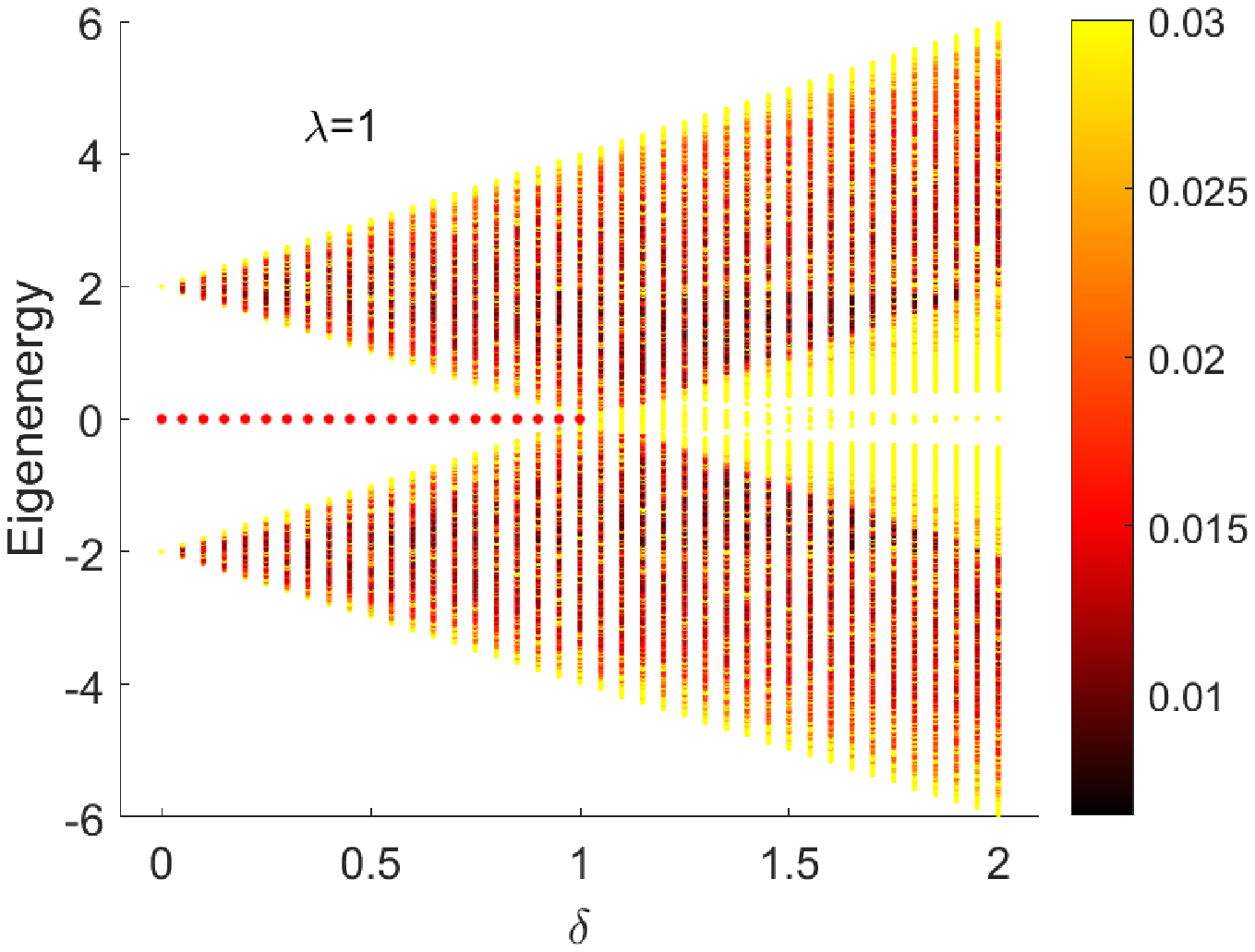}
    \end{minipage}}\\

    \subfigure[]{
    \begin{minipage}[t]{0.5\textwidth}
    \centering
    \includegraphics[width=1\textwidth]{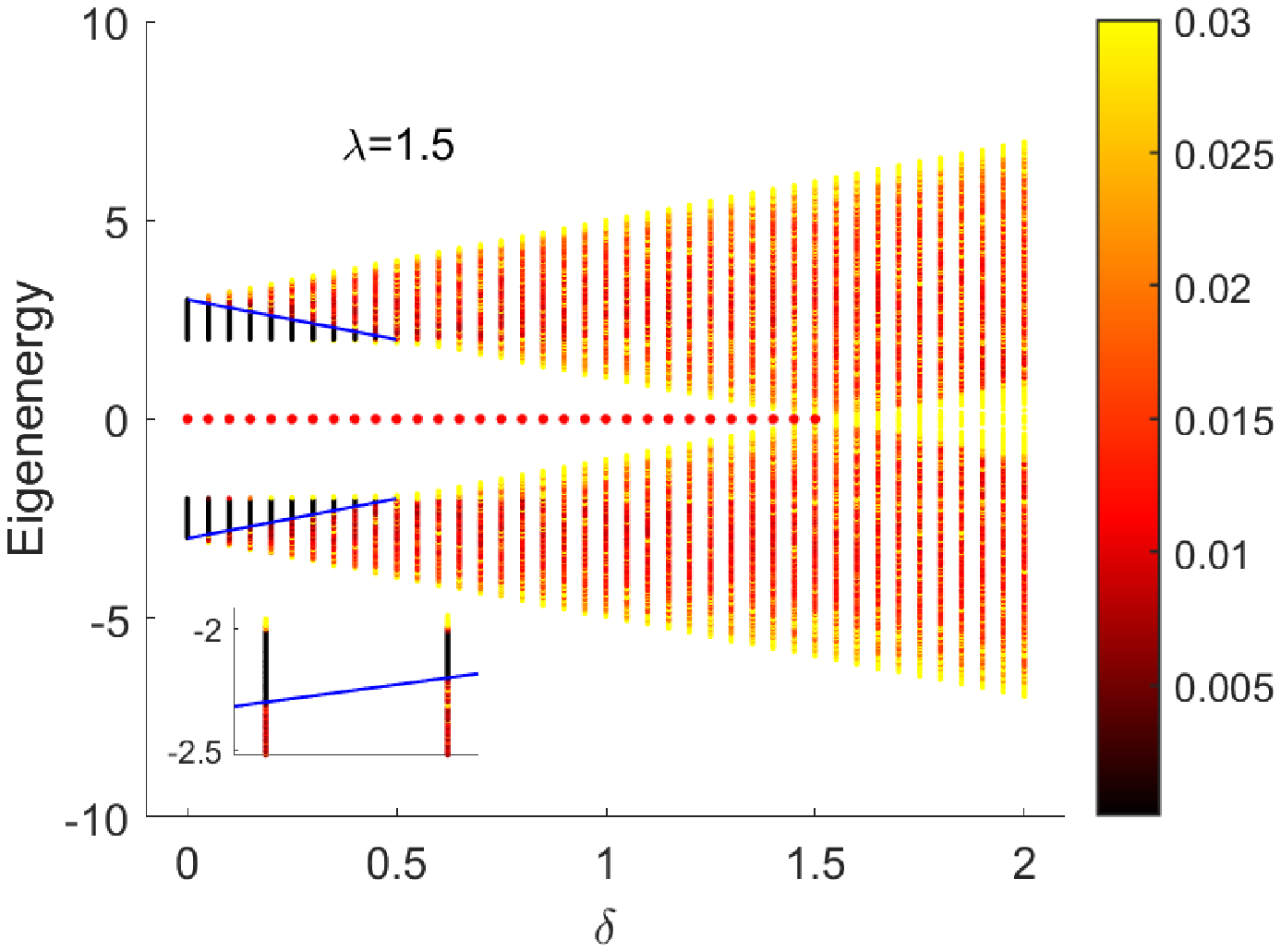}
    \end{minipage}}
    \subfigure[]{
    \begin{minipage}[t]{0.5\textwidth}
    \centering
    \includegraphics[width=1\textwidth]{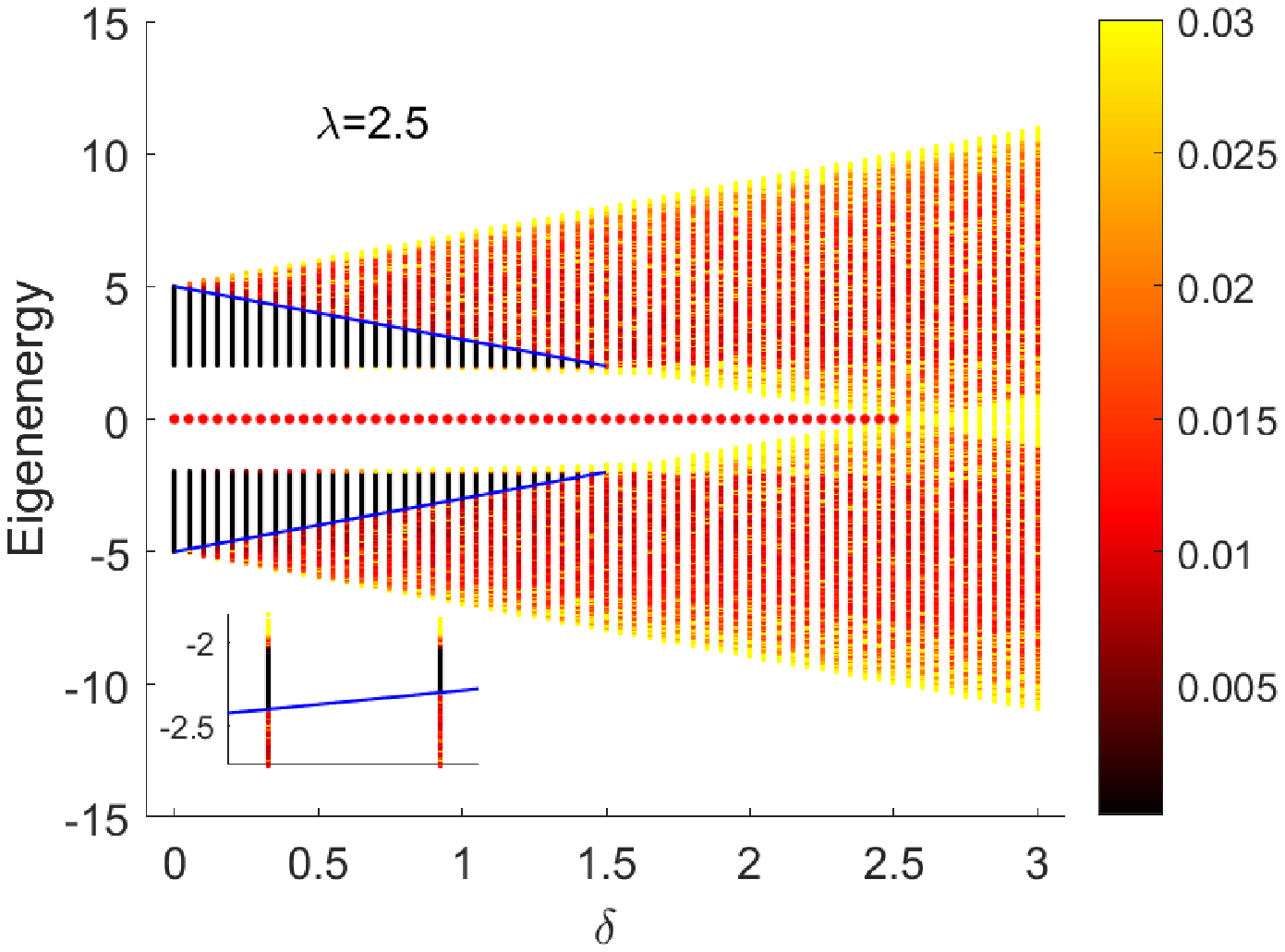}
    \end{minipage}}
    \end{tabular}
\caption{(Color online) Eigenvalues and IPR as a function of the disorder strength $\delta$, with $\lambda=~(a)~0.5,~(b)~1,~(c)~1.5,~(d)~2.5$. The red dots in the midgap represent nontrivial zero-energy modes. The blue solid lines represent the boundary between spatially localized
and extended states, i.e., mobility edges $E_{\pm c}$. In the inset we plot the blow up of eigenvalues near $E=-2$. }
\label{007}
\end{figure*}
Following the similar procedure in the last section about the AA quasiperiodic disorder, we first investigate the location of the gap-closing point with the increasing of $\delta$ for different $\lambda$'s. In Fig.~\ref{005}, we plot the variation of the energy gap $\Delta_g$. It is shown that there are finite gaps as $\delta$ reaches a critical value: $\lambda$. However, unlike the AA quasiperiodic disorder, the energy gap $\Delta_g$ at this critical point still seems to be finite. In fact this is due to the finite-size effect, and we perform the finite-size analysis in
the inset of Fig.~\ref{005}. As shown, the scaling behaviors of $\Delta_g$ when $\lambda=1$ and $\delta=0.8$, $0.9$ vary smoothly, and $\Delta_g$ is obviously finite in the large $L$ limit. Whereas when $\delta=1.1$, $\Delta_g$ vanishes quickly as $L$ increases. The interest situation is associated with the critical point $\delta=1$ where $\Delta_g$ seem to be finite when $L$ is small, whereas the scaling behavior of $\Delta_g$ exhibits a power-law decreasing trend as $L$ increases. We numerically fit the varying curve and obtain the expression $\Delta_g=1.514 L^{-0.3356}$. Hence $\Delta_g$ vanishes at the critical point $\delta=1$ as $L\rightarrow\infty$ , which indicates that $\delta=\lambda$ denotes the gap-closing point. We systematically calculate the energy gap $\Delta_g$ with some other sets of $\delta$ and $\lambda$ and find all gap-closing points indeed stay at $\delta=\lambda$.

We also wonder if there exists a topological invariant in this slowly varying quasiperiodic disordered SSH chain by calculating the total transfer matrix. By implementing the same procedure, we plot the numerical results of $T$ and $R$ of the midmost excitations
for $\lambda=0.5$, $1.5$ in Fig.~\ref{006}. Surprisingly, $T$ always approaches $0$ and $R$ always approaches $1$ with the growth of the disorder strength $\delta$ even when $\delta>\lambda$. This result obviously differs from that by studying the gap-closing point. According to the bulk-boundary correspondence, zero-energy modes are protected by the energy gap, hence they vanish at the gap-closing point which is accompanied with the topological phase transition. To clarify this situation we plot the midmost excitation (the $5001$th eigenvalue) in the inset of Fig.~\ref{006}. It can be found that eigenvalues have almost the same numerical accuracy $10^{-16}$ with the increase of $\delta$, i.e., the slow varying disorder induces trivial zero-energy modes at the chain even if the energy gap closes. The similar phenomenon has been discovered by studying the topological phase transition for a one-dimensional $p$-wave superconductor in slowly varying incommensurate potentials~\cite{liu2}. This is why the transfer matrix breaks down in this model since the validity of it is based on the assumption that the topological phase transition is accompanied with vanishing of zero-energy modes. Thus we have not found a topological invariant to denote the topological phase transition in the presence of the slowly varying quasiperiodic disorder, instead we conjecture that it occurs at the gap-closing point $\delta=\lambda$.

\begin{figure}
  \centering
  \includegraphics[width=0.5\textwidth]{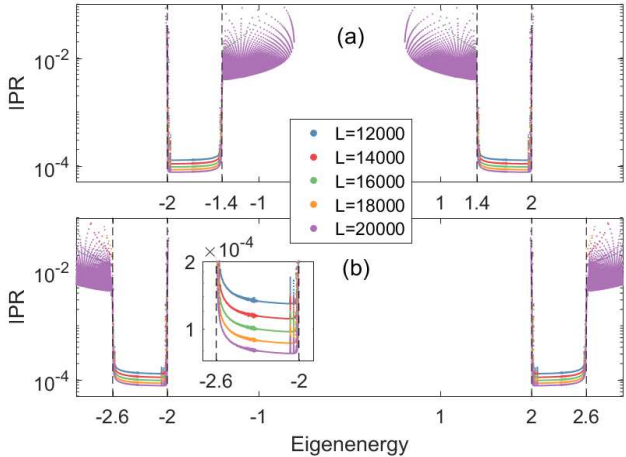}\\
  \caption{(Color online) The finite size analysis of IPR versus eigenvalues for (a)~$(\lambda,\delta) = (0.5,0.2)$, (b)~$(\lambda,\delta) = (1.5,0.2)$. The total number of sites $L$ increases from $12000$ to $20000$. The blow up of the inset shows the scaling behavior of IPR for extended states.}
  \label{008}
\end{figure}
Due to the slowly varying quasiperiodic disorder accompanied with mobility edges~\cite{liu2,liu1,li2017mobility,biddle2010predicted,ganeshan2015nearest}, we wonder the location of mobility edges in the spectrum when
this type of disorder is introduced in 1D SSH chain. To clarify the localized property of this model we present numerical analysis by applying typical diagnostic technique in disordered systems, the inverse participation ratio (IPR). The IPR of a normalized wave function $\psi$ is defined as~\cite{IPR1,IPR2}
\begin{equation}
\text{IPR}^{(n)}=\sum_{i=1}^{L} \left|\psi^{(n)}_{i}\right|^{4},
\end{equation}
where $L$ denotes the total number of sites and $n$ is the energy level index.
It is well known that the IPR of an extended state scales like $L^{-1}$ and approaches
 $0$ in the thermodynamic limit, while it is finite even if $L$ is very large for a localized state.
Figure~\ref{007} plots eigenvalues of the system and IPR's of the corresponding wave-functions as a function of the disorder strength $\delta$ with $\lambda=0.5,~1,~1.5$ and $2.5$.  The red dots in the midgap represent nontrivial zero-energy modes, which are the topological feature of 1D SSH chain. And gap-closing points are exactly located at $\delta=\lambda$ as analyzed in Fig.~\ref{005}. Different colors of the eigenvalue curves indicate different magnitudes of IPR. The black eigenvalue curves denote the extended states with very low magnitude of the IPR (around $10^{-4}$), and the red and yellow eigenvalue curves denote the localized states with high magnitude of IPR ($> 10^{-3}$). The blue solid lines represent two mobility edges $E_{\pm c}$. It is clear that these two boundaries are exactly located between spatially localized and extended states. When $\lambda=0.5$ the explicit expression of mobility edges is $E_{\pm c}=\pm2(\lambda+\delta)$ as shown in Fig.~\ref{007}(a), when $\lambda=1$ there exist no mobility edges as shown in Fig.~\ref{007}(b), when $\lambda=1.5,~2.5$ the explicit expression of mobility edges is $E_{\pm c}=\pm2(\lambda-\delta)$ as shown in Fig.~\ref{007}(c) and (d). Besides $E_{\pm c}$, in the spectrum with the existence of mobility edges, wave-functions with eigenvalues outside the region $[-2,2]$ are also localized as shown in the inset of Fig.~\ref{007}(a) when $\lambda<1$, whereas wave-functions with eigenvalues inside the region $[-2,2]$ are also localized as shown in the inset of Fig.~\ref{007}(c) and (d) when $\lambda>1$. This situation is interesting but hard to give qualitative explanation. The dimerization strength $\lambda=1$ is special, where the intracell hopping $t_1=-\delta_{i}$ and the intercell hopping $t_2= 2+\delta_{i}$. As $\delta$ varies from 0, the spectrum of the system becomes degenerate when $\delta=0$, hence mobility edges can not exist. Therefore the locations of mobility edges depend not only on the disorder strength $\delta$, but also on the dimerization strength $\lambda$.  We also implement calculations for various sets of $(\lambda,\delta)$ to ensure that mobility edges in the spectrum are indeed located at $E_{\pm c}$ and $\pm2$.

To consolidate the results of mobility edges, we plot IPR of the corresponding wave-functions as a function of eigenvalues for various $L$'s in Fig.~\ref{008}. As the eigenvalue varies, we find that IPR changes dramatically from the order of magnitude $10^{-2}$ (a typical value for the localized states) to $10^{-4}$ (a typical value for the extended states), or inversely at certain energies which exactly correspond to $E_{\pm c}$ and $\pm2$ as indicated by previous analysis. With the increase of $L$, the magnitude of IPR of extended states becomes lower as shown in the inset, hence the change of IPR at these turning points becomes more abrupt. Thus, we deduce that there will be a jump at these turning points in the thermodynamic limit $L\rightarrow\infty$.
This jumping phenomenon of IPR indeed indicates that there exist mobility edges at $E_{\pm c}$ and $\pm2$ in the energy spectrum.

\section{Conclusions}
\label{n4}
In this work we have studied the topological phase transition in 1D SSH chain subject to two types of quasiperiodic hopping disorder, one is the AA quasiperiodic disorder $\delta_{i}=\delta\cos(2\pi\beta{i})$ and the other is the slowly varying quasiperiodic disorder $\delta_{i}=\delta\cos(2\pi\beta{i^{v}})$ with $0<v<1$, we find following interesting features of two models.

(1) When $\delta_{i}=\delta\cos(2\pi\beta{i})$, the gap-closing point is located at the critical disorder strength $\delta=1+\lambda$. We also calculate the transmission probability and the reflection probability by applying the transfer matrix method, which clearly demonstrates the topological phase transition indeed occurs at $\delta=1+\lambda$, i.e. the gap-closing point.

(2) When $\delta_{i}=\delta\cos(2\pi\beta{i^{v}})$, the gap-closing point is located at the other critical disorder strength $\delta=\lambda$. However, the transfer matrix breaks down in this model, and we conjecture that the topological phase transition
occurs at the location of the gap-closing point. We further investigate the localized property of this model, and identify that there exist mobility edges in the spectrum when the dimerization strength is unequal to 1.

In general the random disorder will shift the separation point between the topologically trivial and non-trivial phases of the translation-invariant system into a finite region under the finite size, and we may get a smooth curve instead of a step function with respect to the topological phase transition in the parameter space.
However, by varying the parameters of these models the sharp topological phase transitions emerge. We believe that these interesting features and open questions (such as the convergence of Eq.~(\ref{Z1})) will bring new perspectives to a
wide range of topological and disordered systems.

\begin{acknowledgments}
G. H. acknowledges support from the National Natural Science Foundation of China under Grant No.~11674051.

\end{acknowledgments}

\bibliographystyle{apsrev}
\bibliography{Review}

\end{document}